\documentclass[pra,onecolumn,floatfix,a4paper,superscriptaddress]{revtex4}
\usepackage{bm,color,graphicx,amsmath,txfonts}

\usepackage[colorlinks, citecolor=blue,linkcolor=blue]{hyperref}

\newcommand{\Tr}{{\rm Tr}}

\newcommand{\e}{{e}}

\begin{document}

\title{Quantum Thermodynamics and Hierarchy of Quantum Correlations and Fidelity of Teleportation in a Two Coupled Double Quantum Dots}

\author{M. Amazioug} \thanks{amazioug@gmail.com}
\affiliation{LPTHE-Department of Physics, Faculty of sciences, Ibnou Zohr University, Agadir, Morocco.}

\author{M. Daoud}  \thanks{m\_daoud@hotmail.com}
\affiliation{LPMS, Department of Physics, Faculty of Sciences, University Ibn Tofail, K\'enitra, Morocco.}

\begin{abstract}

We explore the quantum correlations, fidelity and quantum thermodynamics of two coupled double quantum dots containing two excess electrons. In this regard, we investigate and compare the evolution of those measures under thermal effects and tunneling coupling. We find the hierarchy of quantum correlations, and one-way steering between the two quantum dots. We found, as expect that the quantum correlations are diminishes by increasing the values of temperature. We show that this state can be used for quantum teleportation. On the other, we address the extracting work and efficiency of the state. We compare the extraction work with the bare energies. Our results show that quantum dots states have a reliable and better capacity to preserve quantum correlations and remain one of the good resources for the deployment of quantum information processing protocols.

\end{abstract}

\date{\today}

\maketitle

\section{Introduction}

In recent years, numerous works have been devoted to the theoretical research in the field of quantum thermodynamics \cite{SVinjanampathy2016}. The quantum thermodynamics and its applications is one of the most intriguing issues in this new topic of research on quantum information \cite{HTQuan2007,HTQuan2006,SWKim2011,TEHumphrey2002,BLeggio2015,MJames2016,AHewgill2018}.  In particular, extracting work is a possible quantum process that has been considered theoretically in several works \cite{MHorodecki2013,PSkrzypczyk2014}. Experimentally, this is difficult to achieve. Indeed, numerous efforts have been made to discover how to develop useful thermodynamic protocols \cite{MLostaglio2018,CPerry2018}.

In 1935, Einstein, Podolsky and Rosen (EPR) presented the EPR paradox, illustrating the observation of a spooky action-at-a-distance between two quantum component systems \cite{EPR1935}. This paradox serves as an example of quantum mechanics' nonlocality. The EPR paradox was addressed by Schrodinger, who was the first physicist to suggest entanglement and quantum steering \cite{Schrodinger1935}. Bell demonstrates that the theory of local hidden variables cannot predict all quantum correlations shared between spatially separated systems by local quantum measurements \cite{JSBell1964}. The global states of two or more quantum systems that are entangled cannot be divided into the direct products of individual subsystems. This means that the behavior of one entangled system is intimately connected to the behavior of the other, regardless of the distance between them. This phenomenon is at the heart of numerous useful applications in the context of quantum information processing and quantum computing. Entanglement is not only a very significant phenomenon but also carries fascinating quantum mechanical properties \cite{CHBennett1993,CHBennett1996,LAmico2008}. Moreover, There are states that behave in a non-local state while remaining separable. Quantum Discord was developed in order to measure such non-local correlations. It is a quantifier of the total non-local correlations of a quantum system. Besides, if we consider Alice and Bob to have a shared entangled pair, Alice can influence Bob's state by making measurements only on her side of the system. Quantum steering, or EPR steering, is the term used to describe this kind of quantum correlation. Additionally, it is a tool for quantum teleportation \cite{YFan2022}. The Bell non-locality and entanglement are on either side of it. In this order, quantum steering vs entanglement and extracting work in an anisotropic two-qubit Heisenberg model in presence of external magnetic fields with DM and KSEA Interactions, was investigated \cite{amaziougqubit,Daoud23}. Recently, the quantum correlations and fidelity are taking into consideration in the differents works \cite{Daoud18,Eleuch22,MNIbrahim23,MYAbdRabbou22,urRahman23,JLalita23}.

Recently, due to the fact that semiconductor nanostructures like quantum dots \cite{JRPeta2005,DPress2008} and double quantum dots (also referred to as quantum dot molecules) \cite{GShinkai2004,DGAusting1998} are promising candidates for the physical implementation of quantum information processing, solid-state realizations of them have drawn a lot of attention \cite{ESophia2012}. The chemists
Moungi G. Bawendi, Louis E. Brus and Alexey I. Yekimov won the 2023 Nobel Prize "for the discovery and synthesis of quantum dots". According to some proposals, quantum dots could use both spin \cite{MBenito2017,DLoss1998,BDanjou2019} and charge \cite{JGorman2005} simultaneously \cite{ZShi2012,YCYang2020} as qubits. Additionally, these quantum systems have paramount importance for their promising capabilities to incorporate into modern electronics \cite{TItakura2003,MUrdampilleta2015}. It was described how to coherently control tunneling in an asymmetric double quantum dot \cite{JMVillasBoas2004}.

In this paper, we investigate the nonclassical correlations through differents measures such as quantum discord, concurrence, steerabilities and Bell non-locality, in a double quantum dots system. We employ quantum discord to quantify the amount of quantum correlations beyond entanglement which is quantified by the concurrence.
We use quantum steering to characterize the steerabilities of the two-quantum dots. We prove the hierarchy of quantum correlations under thermal effect for various values of coupling tunneling, i.e., Our results show that the quantum discord is strong for a wide range of temperature. Also, the extracted work and efficiency are quantified under thermal effect for various values of coupling tunneling. Besides, the extracted work and bar energies are compared under the impact of temperature.

The article is organized as follows. In Section II, we study the model and we prove the density matrix of double quantum dots. We quantify and we compare diffrents the quantum correlations such as quantum discord, concurrence, steerabilities and Bell non-locality in Section III. The extracted work, efficiency and energies are devoted in Section IV. Concluding remarks close this paper.

\section{Model}

The system under consideration consists of two sets of double quantum dots (DQDs) coupled via the strength of the tunneling coupling $\Delta$, each filled with a single electron, with an additional electron present in either the left dot ($\left|L\right>$) or the right dot ($\left|R\right>$). The pseudospin Hamiltonian of the DQDs, writes as \cite{FFFanchini2010}
\begin{equation}
\begin{array}{ccc}
\mathbf{H} & = & \Delta (\sigma_{1}^{x}\otimes\mathbf{1}+\mathbf{1}\otimes\sigma_{2}^{x})+J\sigma_{1}^{z}\otimes\sigma_{2}^{z}
\end{array}
\label{H1}
\end{equation}
where $J$ is the Coulomb interactions between the electrons, which favors anti-parallel configurations $\left|LR\right>$ and $\left|RL\right>$ over the parallel ones $\left|LL\right>$ and $\left|RR\right>$, while $\sigma_{1(2)}^{i}(i=x,z)$ are the Pauli matrices. We make into account the convention $\left|0\right>\equiv\left|L\right>$ and $\left|1\right>\equiv\left|R\right>$ to show the electron that is present in either the left dot, $\left(L\right)$, or the right dot, $\left(R\right)$. The Hamiltonian (\ref{H1}) in the basis $\left\{ \left|LL\right>,\left|LR\right>,\left|RL\right>,\left|RR\right>\right\} $, writes as
\begin{equation}
   \label{eq:HXzz}
   {\cal H}=
	 \left(
      \begin{array}{cccc}
      J & \Delta & \Delta & 0\\
      \Delta & -J & 0 & \Delta\\
      \Delta & 0 & -J & \Delta\\
      0 & \Delta & \Delta & J
      \end{array}
   \right),
\end{equation}
The corresponding eigenvalues  
\begin{eqnarray}
\mathcal{E}_1 &=& -J \quad,\quad \mathcal{E}_2 = J \quad,\quad \mathcal{E}_3 = -\sqrt{J^2 + 4\Delta^2}\quad,\quad \mathcal{E}_4 = \sqrt{J^2 + 4\Delta^2},\label{eq:1a} 
\end{eqnarray}
The corresponding eigenstates 
\begin{eqnarray}
|\phi_1\rangle &=& -\frac{1}{\sqrt{2}} |LR\rangle+\frac{1}{\sqrt{2}}|RL\rangle\quad,\quad |\psi_2\rangle = -\frac{1}{\sqrt{2}}|LL\rangle+\frac{1}{\sqrt{2}}|RR\rangle,\nonumber \\ 
&&|\psi_3\rangle = \beta_+(|LL\rangle-\alpha_+|LR\rangle - \alpha_+|RL\rangle+|RR\rangle)\quad,\quad |\psi_4\rangle = \beta_-(|LL\rangle-\alpha_-|LR\rangle - \alpha_-|RL\rangle+|RR\rangle),\label{eq:1a} 
\end{eqnarray}
where $\beta_\pm = \frac{1}{\sqrt{2(1+\alpha^2_\pm)}}$ and $\alpha_\pm = \frac{J\pm \sqrt{J^2+4\Delta^2}}{2\Delta}$. The density operator of the system, writes as
\begin{equation} \label{rhozz}
\varrho(T)= \frac{\e^{-\beta \mathcal{H}}}{Z} = \frac{1}{Z}\sum^4_{j=1} \e^{-\beta \mathcal{E}_j} |\phi_j\rangle \langle \phi_j|,
\end{equation}
where $Z = {\rm Tr }[\e^{-\beta \mathcal{H}}]$ is the partition function of the system and $\beta=1/(k_B T)=1/T$ ($k_B$ is the Boltzmann constant). From the equation (\ref{rhozz}), the thermal state $\varrho(T)$ can be cast in the matrix form
\begin{equation}
   \label{rhozz2}
   {\varrho}(T)=
	 \left(
      \begin{array}{cccc}
      \varrho_{11}& \varrho_{12} & \varrho_{12} & \varrho_{14}\\
      \varrho_{12} & \varrho_{22} & \varrho_{23}& \varrho_{12}\\
      \varrho_{12} & \varrho_{23} & \varrho_{22} & \varrho_{12}\\
      \varrho_{14} & \varrho_{12} & \varrho_{12} & \varrho_{11}
      \end{array}
   \right)
\end{equation}
where 
\begin{equation}
\varrho_{11} = \bigg(\frac{-\e^{\beta\mathcal{E}_2}}{2}+\beta_+^2\e^{-\beta\mathcal{E}_3}++\beta_-^2\e^{-\beta\mathcal{E}_4}\bigg)/Z\quad,\quad \varrho_{22} = \bigg(\frac{-\e^{\beta\mathcal{E}_1}}{2}+\alpha^2_+\beta_+^2\e^{-\beta\mathcal{E}_3}+\alpha^2_-\beta_-^2\e^{-\beta\mathcal{E}_4}\bigg)/Z,
\end{equation}

\begin{equation}
\varrho_{12} = -\bigg(\alpha_+\beta_+^2\e^{-\beta\mathcal{E}_3}+\alpha_-\beta_-^2\e^{-\beta\mathcal{E}_4}\bigg)/Z\quad,\quad \varrho_{14} = \bigg(-\frac{-\e^{\beta\mathcal{E}_2}}{2}+\beta_+^2\e^{-\beta\mathcal{E}_3}++\beta_-^2\e^{-\beta\mathcal{E}_4}\bigg)/Z,
\end{equation}

\begin{equation}
\varrho_{23} = \bigg(-\frac{-\e^{\beta\mathcal{E}_1}}{2}+\alpha^2_+\beta_+^2\e^{-\beta\mathcal{E}_3}+\alpha^2_-\beta_-^2\e^{-\beta\mathcal{E}_4}\bigg)/Z,
\end{equation}
where $Z = 2(\cosh{(J\beta)} + \cosh{(\beta \sqrt{J^2 + 4\Delta^2})})$.

\section{Quantum correlation quantifiers}

In this section, we will go over the quantum correlations contained in the two quantum dots system described by the $4\otimes 4$ density matrix (\ref{rhozz2}) and compute some quantifiers, such as quantum discord, concurrence, quantum steering and Bell non-locality.

\subsection{Bell non-locality}\label{Bell}

In this subsection, we will quantify the violation of the CHSH inequality. For two-qubit systems, the mixed state (\ref{rhozz2}) can be written with Fano–Bloch decomposition as \cite{FBlock1946,UFano1983}
\begin{equation}  \label{FB}
\varrho = \frac{1}{4}\bigg[\mathbf{1}\otimes \mathbf{1}+\sum^{3}_{j=1}\bigg(u_j\sigma^1_j\otimes \mathbf{1}+v_j\mathbf{1}\otimes\sigma_j^2\bigg)+\sum_{j,k}^3 \mathbb{R}_{jk}\sigma^1_j\sigma^2_k\bigg],
\end{equation}
where $\mathbf{1}$ is the identity operator, with $u_j$, $v_j$ and $\mathbb{R}_{jk}$ represent the real parameters, while the Pauli matrices denote by $\sigma_{j = 1,2,3}$. Then, a correlation matrix writes as
\begin{equation}
   \label{Z}
   {\cal R}=
	 \left(
      \begin{array}{cccc}
      1 & v^T \\
      u & \mathbb{R} 
      \end{array}
   \right),
\end{equation}
with $u = (u_1,u_2,u_3)^T$, $v^T = (v_1,v_2,v_3)$ and a $3 \times 3$ matrix with $\mathbb{R}_{jk}$-elements is represented by $\mathbb{Z}$. The elements of the matrix are given by ${\cal R}_{jk} = \text{Tr}(\varrho\sigma^1_j \otimes \sigma_k^2)$ for $j, k = 0,1,2,3$ with $\sigma_0 = \mathbf{1}$. The CHSH inequality is violated if and only if $\mathcal{M(\varrho)}>1$ \cite{Horodecki1996}, where $\mathcal{M(\varrho)}=\max_{j<k}[\lambda_j+\lambda_k]\leq 2$, and $\lambda_j$ $(j = 1; 2; 3)$ are the eigenvalues of the matrix $\mathcal{U}=\mathbb{R}^T\mathbb{R}$. According to the Horodecki criterion \cite{Horodecki1996}, $B = 2\sqrt{\mathcal{M(\varrho)}}$, where $\mathcal{M(\varrho)}=\max_{j<k}[\lambda_j+\lambda_k]$ with $j, k=1, 2, 3$. In order to quantify the violation of the Bell inequality, one can use $\mathcal{M(\varrho)}$ or, equivalently,
\begin{equation}
\mathcal{B} = \sqrt{\max[0,\mathcal{M(\varrho)}-1]},
\end{equation}
which yields $\mathcal{B} = 0$ if the Bell inequality is not violated and $\mathcal{B} = 1$ for its maximal violation.

\subsection{Quantum steering}\label{S1}

The degree of steerability from quantum dot 1 (Alice: d1) to quantum dot 2 (Bob: d2). It writes as follows \cite{DCavalcanti2016, AbdRabbou2022}
\begin{equation}
\mathbb{S}^{d1\to d2}=\max\bigg[0,\frac{\mathcal{L}_{d1d2}-2}{\mathcal{L}_{\max}-2}\bigg],
\end{equation}
and from quantum dot 2 to quantum dot 1
\begin{equation}
\mathbb{S}^{d2\to d1}=\max\bigg[0,\frac{\mathcal{L}_{d2d1}-2}{\mathcal{L}_{\max}-2}\bigg],
\end{equation}
where $\mathcal{L}_{\max}=6$, when the system was prepared in Bell
states. The expression of the EPR inequality of steerabilities from $d1$ to $d2$, given by
\begin{equation} \label{Ld1d2}
\mathcal{L}_{d1d2}=H(\sigma_x^{d2}|\sigma_x^{d1})+H(\sigma_y^{d2}|\sigma_y^{d1})+H(\sigma_z^{d2}|\sigma_z^{d1})\geq 2
\end{equation}
where $\sigma_x,\sigma_y$ and $\sigma_z$ represent the Pauli spin  operator as measurements and the symbol $H$ stands for the Shannon entropy and $H(d2|d1) = H(\varrho_{d1d2}) - H(\varrho_{d1})$ represents the conditional Shannon entropy. Then, the explicit expression of the quantity (\ref{Ld1d2}), writes as 
\begin{equation}\label{e3}
	\begin{split}
	\mathcal{L}_{d1d2}=&\frac{1}{2}\bigg\{2\varrho_{11}Log(2\varrho_{11})+4\varrho_{11}Log(4\varrho_{11})+8\varrho_{22}Log(4\varrho_{22})+[1-2(4\varrho_{12}-\varrho_{14}-\varrho_{23})]Log[1-2(4\varrho_{12}-\varrho_{14}-\varrho_{23})]
\\& +2[1-2(-\varrho_{14}+\varrho_{23})]Log[1-2(-\varrho_{14}+\varrho_{23})]+2[1+2(-\varrho_{14}+\varrho_{23})]Log[1+2(-\varrho_{14}+\varrho_{23})]
\\& +2[1-2(\varrho_{14}+\varrho_{23})]Log[1-2(\varrho_{14}+\varrho_{23})]+[1+2(4\varrho_{12}+\varrho_{14}+\varrho_{23})]Log[1+2(4\varrho_{12}+\varrho_{14}+\varrho_{23})]\bigg\}
	\end{split}
\end{equation}

and $\mathcal{L}_{d2d1}$ quantifies the steering from $d2$ to $d1$. It is given by
\begin{equation}\label{e3}
	\begin{split}
	\mathcal{L}_{d2d1}=&\frac{1}{2}\bigg\{2\varrho_{11}Log(2\varrho_{11})+4\varrho_{11}Log(4\varrho_{11})+8\varrho_{22}Log(4\varrho_{22})+2[1-2(-\varrho_{14}+\varrho_{23})]Log[1-2(-\varrho_{14}+\varrho_{23})] 
	\\&+2[1+2(-\varrho_{14}+\varrho_{23})]Log[1+2(-\varrho_{14}+\varrho_{23})]
 +2[1-2(\varrho_{14}+\varrho_{23})]Log[1-2(\varrho_{14}+\varrho_{23})]
 \\&+[1+2(4\varrho_{12}+\varrho_{14}+\varrho_{23})]Log[1+2(4\varrho_{12}+\varrho_{14}+\varrho_{23})]\bigg\}
	\end{split}
\end{equation}

\subsection{Quantum entanglement}\label{EoF}

Now we discuss the entanglement behavior the two quantum dots by employing the concurrence \cite{WKWootters1998}. The expression of the concurrence is characterized for the bipartite state, by the quantity
\begin{equation}
\mathcal{C}(\varrho) = \text{max}\bigg[0,\sqrt{\Lambda_4}-\sum^3_{j=1}\sqrt{\Lambda_j}\bigg].
\end{equation}
Here $\Lambda_i$ ($i=1,2,3,4$), are non-negative eigenvalues of the matrix $\tilde\varrho$
\begin{equation}
\tilde\varrho=\varrho(\sigma_y \otimes \sigma_y)\varrho^\star (\sigma_y \otimes \sigma_y)
\end{equation}
where $\Lambda_4>\Lambda_2>\Lambda_1>\Lambda_3$, $\sigma^\star$ is the complex conjugation of a matrix $\varrho$ and $\sigma_y = i(-| 0\rangle\langle 1|+| 1\rangle\langle 0|)$. It is important to note that the concurrence $\mathcal{C}(\varrho)$ ranges from 0 to 1: (*) $\mathcal{C}=0$, indicates that the state quantum system is separable, (**) $\mathcal{C}>0$, the quantum system is not separable, (***) $\mathcal{C}=1$, the quantum system in a maximum entangled state.

\subsection{Quantum discord}

In the Bloch normal form, the density matrix for two-quantum dot state, Eq. (\ref{FB}), can be expressed as a linear combination of Pauli matrices by performing local unitary transformations \cite{SLuo2008,FVerstraete2001,SLuo08}
\begin{equation}  \label{FB2}
\varrho = \frac{1}{4}\bigg[\mathbf{1}\otimes\mathbf{1}+\sum^{3}_{j=1}\bigg(A_j\sigma^1_j\otimes \mathbf{1}+B_j\mathbf{1}\otimes\sigma_j^2\bigg)+\sum_{j=1}^3 C_{j}\sigma^1_j\sigma^2_j\bigg].
\end{equation}
We will only consider the states with maximally mixed marginals, which corresponds to a simplified family of states as follows:
\begin{equation}  \label{FB3}
\varrho = \frac{1}{4}\bigg[\mathbf{1}+\sum_{j=1}^3 C_{j}\sigma^1_j\otimes\sigma^2_j\bigg],
\end{equation}
where $C_{j}=\text{Tr}[\varrho (\sigma^1_j\otimes\sigma^2_j)]$ are real. The quantum discord $\mathcal{D}$ can be described as \cite{SLuo08,Adesso11}

\begin{equation}  \label{FB3}
\mathcal{D}(\varrho) = \mathcal{I}(\varrho) - \mathbb{C}(\varrho),
\end{equation}
where $\mathcal{I}$ is the quantum mutual information in $\varrho$ is defined as
\begin{equation}  \label{FB3}
\mathcal{I} = 2+\sum_{j}^3 \delta_{j}\text{Log}(\delta_{j}),
\end{equation}
with $\delta_{j}\in [0,1]$ $(j=0,1,2,3)$, are eigenvalue of $\varrho$:
$\delta_0=(1-C_1-C_2-C_3-C_4)/4\quad;\quad \delta_1=(1-C_1+C_2+C_3+C_4)/4\quad;\quad\delta_2=(1+C_1+C_2-C_3+C_4)/4\quad\text{and}\quad\delta_3=(1+C_1+C_2+C_3-C_4)/4$. The classical correlations $\mathbb{C}$, writes as
\begin{equation}  \label{FB3}
\mathbb{C} = \frac{1-C}{2}\text{Log}\bigg(\frac{1-C}{2}\bigg) + \frac{1+C}{2}\text{Log}\bigg(\frac{1+C}{2}\bigg),
\end{equation}
where $C = \text{max}(|C_1|,|C_2|,|C_3|)$.

\subsection{Results discussions}

\begin{figure}[!htb]
\includegraphics[width=0.4\linewidth]{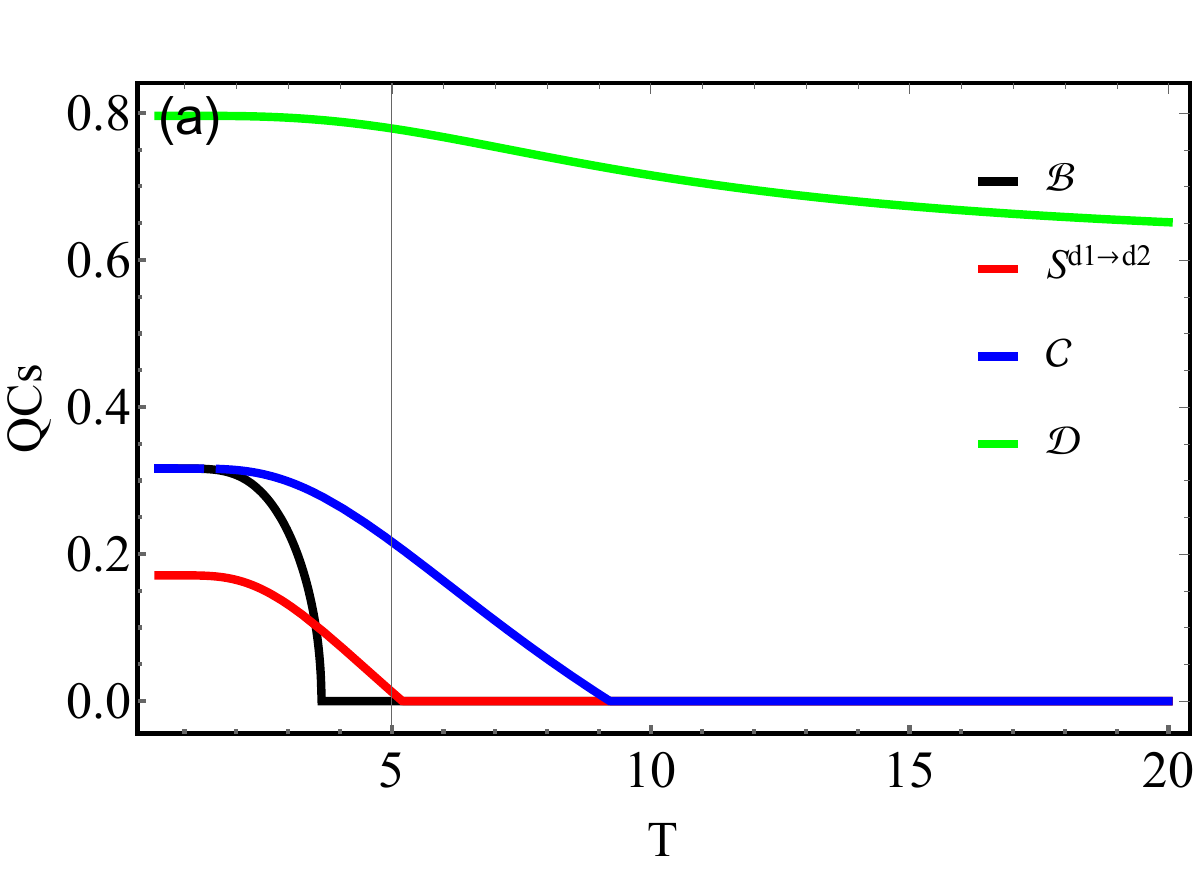}
\includegraphics[width=0.4\linewidth]{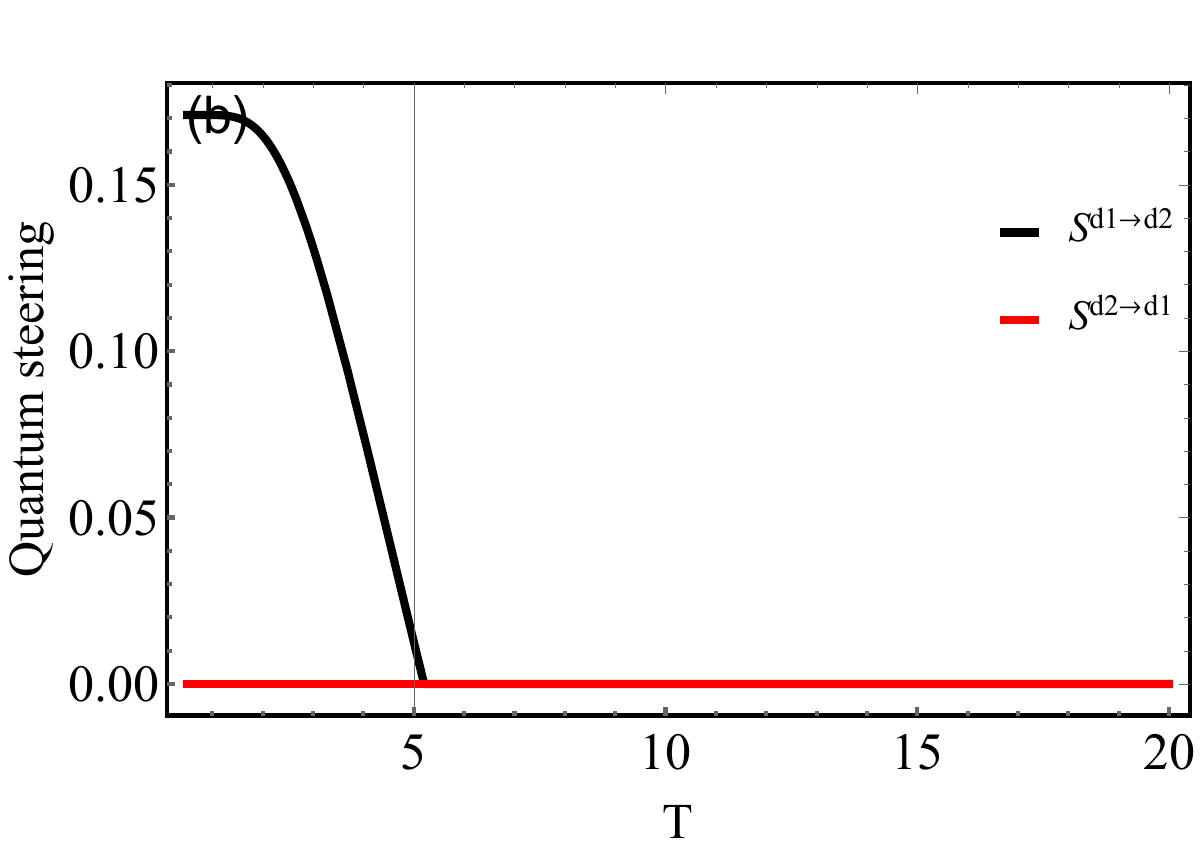}
\caption{Plot of the geometric measure of quantum discord, concurrence, quantum steering, and Bell nonlocality versus the temperature $T$, for $\Delta=9$ and $J=6$.}
\label{QCs}
\end{figure}

In Fig. \ref{QCs}(a), we discuss in steady state the quantum correlations (QCS) such as quantum discord, concurrence, quantum steering, and Bell nonlocality, in a double quantum dots system under thermal noises. This figure, exhibits the hierarchy of quantum correlations. However, the quantum correlations decrease quickly to vanish. While for high temperature, i.e., $6<T<10$, the $\mathcal{D}$ and $\mathcal{C}$ are strong (are not vanishing), as depicted in Fig. \ref{QCs}(a). The state of the two quantum dots has CHSH non-locality when $0<\mathcal{B}\leq 1$, and is also non-separable $\mathcal{C}>0$ (or entangled). The region of CHSH non-locality is small than region of EPR-steering (quantum steering). Besides, the quantum correlations beyond entanglement take place between the two quantum dots, because $\mathcal{D}> 0$ and concurrence is zero ($\mathcal{C}=0$). We remark also, that the quantum steering $\mathbb{S}^{d1\to d2}$ is bounded by the entanglement $\mathcal{C}$. We observe, that quantum steering between the two quantum dots is one-way steering, i.e., $\mathbb{S}^{d1\to d2}>0$ and $\mathbb{S}^{d2\to d1}=0$ (quantum dot 1 can steer quantum dot 2, while quantum dot 2 can not steer quantum dot 1 even if $\mathcal{C}>0$), as depicted in Fig. \ref{QCs}(b). Moreover, the quantum steering between the two quantum dots is no-way steering, is reached for $T\geq 6$, i.e., $\mathbb{S}^{d1\to d2}=\mathbb{S}^{d2\to d1}=0$. Indeed, entangle state is always steerable, while steerable state is not always entangle state.

\section{Maximal fidelity and fidelity deviation}

Quantum teleportation is an essential protocol to transmit quantum information through shared entanglement, local operations, and classical communication \cite{CHBennett1993}. The maximal average fidelity $\mathcal{F(\varrho)}$ in a quantum state $\varrho$ \cite{Horodecki1996,PBadziag2000}, and fidelity deviation $\Delta(\varrho)$ \cite{JBang2018,AGhosal2020} are usually used to determine the quality of a teleportation protocol. The maximum average fidelity (or maximal fidelity) corresponds to the greatest average fidelity achievable by employing local unitary operations and the standard protocol in a strategy. For double quantum dots states with $\text{Det}(\mathbb{R}) < 0$ (where $\mathbb{R}$ is the correlation matrix in Eq. (11)), the maximal average fidelity is quantifiable as \cite{Horodecki1996,AGhosal2020}
\begin{equation}
\mathcal{F(\varrho)} = \frac{1}{6}\bigg(3+\sum^3_{j=1}|r_j|\bigg),
\end{equation}
where $r_j$’s are the eigenvalues of the matrix $\mathbb{R}$. The fidelity deviation is characterized by the standard deviation of fidelity values for every potential input state. For a two-quantum dots state with $\text{Det}(\mathbb{R}) < 0$, the fidelity deviation that corresponds to the ideal protocol is quantified as \cite{JBang2018,AGhosal2020}
\begin{equation}
\Delta(\varrho) = \frac{1}{3\sqrt{10}}\sqrt{\sum^3_{j,k}\bigg(|r_j|-|r_k|\bigg)^2},
\end{equation}

\begin{figure}[!htb]
\minipage{0.32\textwidth}
  \includegraphics[width=\linewidth]{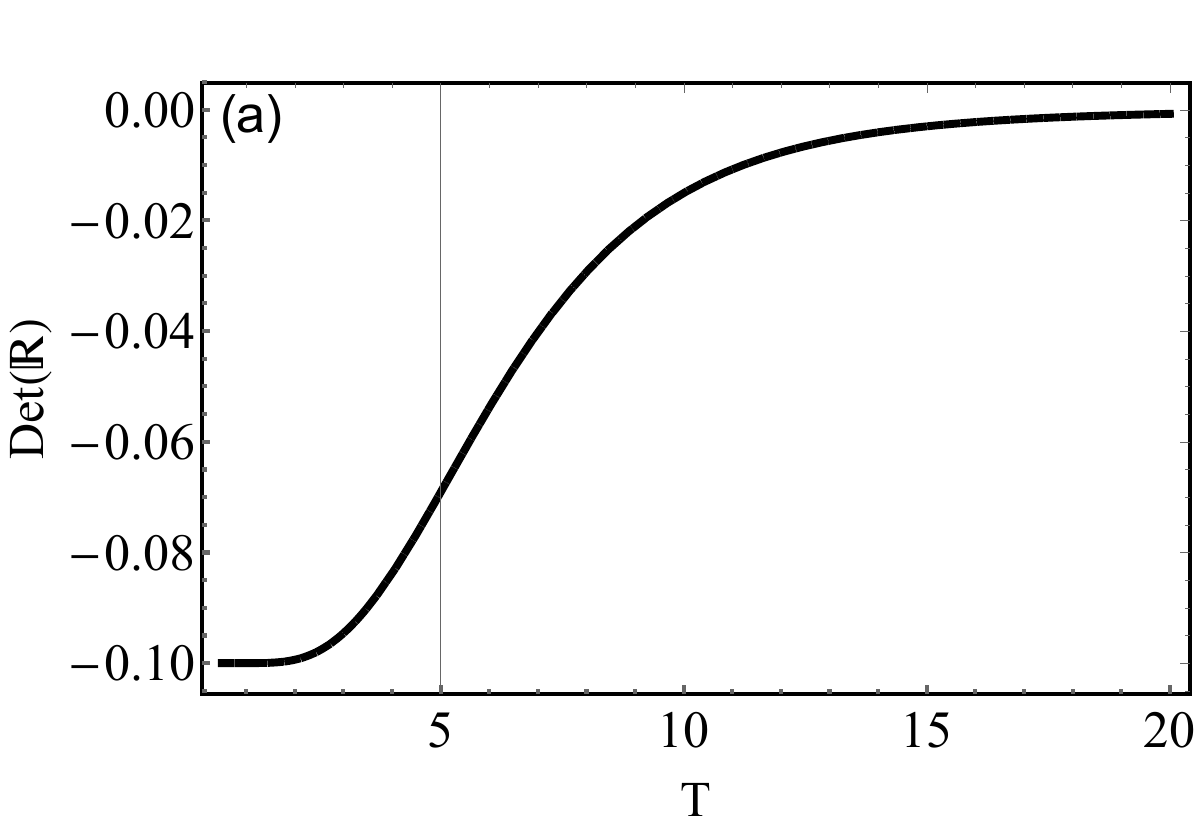}
\endminipage\hfill
\minipage{0.32\textwidth}
  \includegraphics[width=\linewidth]{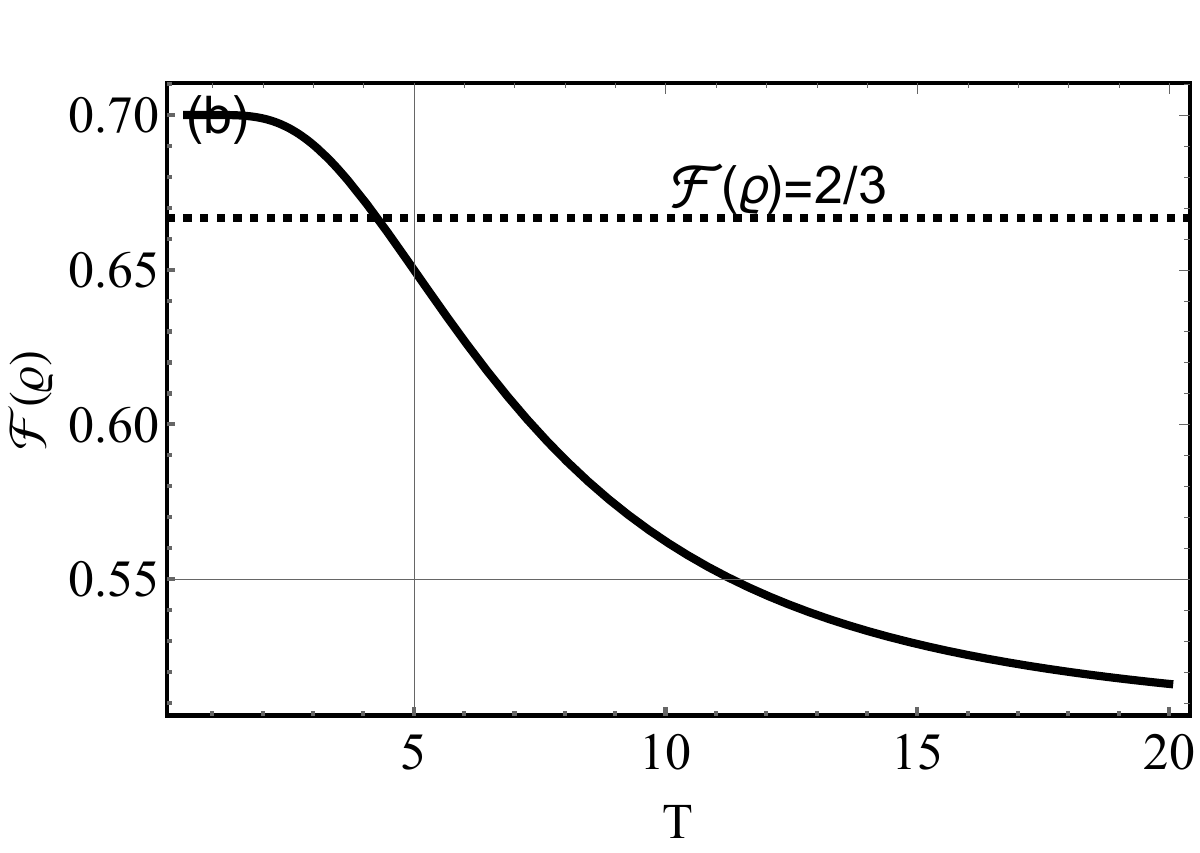}
\endminipage\hfill
\minipage{0.32\textwidth}
  \includegraphics[width=\linewidth]{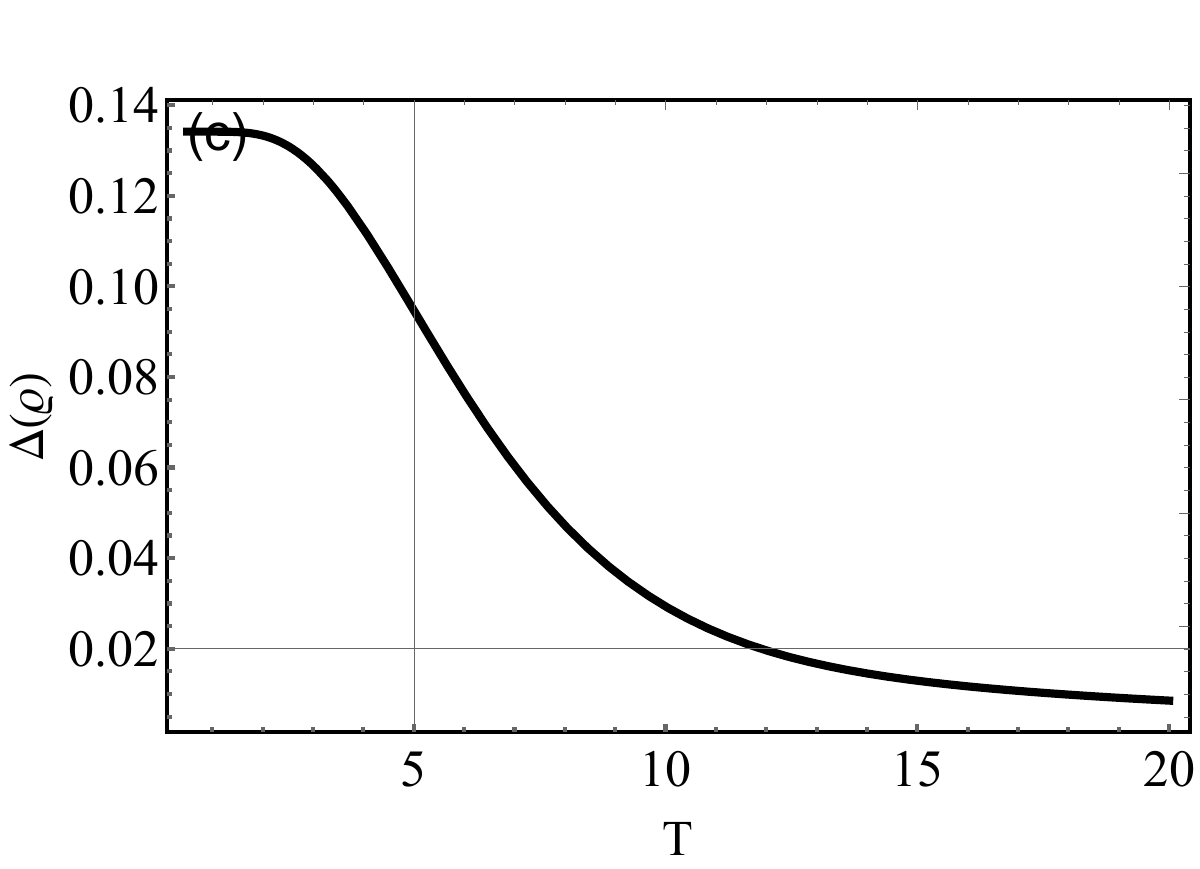}
\endminipage\hfill
\caption{Plot of the $Det(\mathbb{R})$ (a) the Maximal fidelity $\mathcal{F(\varrho)}$ (b) and the fidelity deviation $\Delta(\varrho)$ (c) of the two quantum dots versus the Temperature $T$, with $\Delta=9$ and $J=6$.}
\label{F}
\end{figure}

From Fig. \ref{F}(a), it is clear that the $\text{Det}(\mathbb{R})$, remains negative versus the temperature $T$. In Fig. \ref{F}(b), the maximal fidelity (the average fidelity) $\mathcal{F(\varrho)}$ decreases quickly from its maximum value, under thermal effects. This figure exhibits that the two-quantum dots state is useful for quantum teleportation \cite{AGhosal2020}, i.e., $\mathcal{F(\varrho)}>2/3$, thus $\text{Det}(\mathbb{R})<0$; while the highest average fidelity possible with classical protocols is 2/3. We know that the most desirable states for quantum teleportation are those with zero fidelity deviation, while in our case, we have obtained that the fidelity deviation $\Delta(\varrho)$ is very small (non-zero), even if $\mathcal{F(\varrho)}>2/3$, as depicted in Fig. \ref{F}(b). Also, the quantum correlations, quantum discord, entanglement, EPR-steering and CHSH-non-locality are non vanishing when $\mathcal{F(\varrho)}>2/3$. Besides, for $\mathcal{F(\varrho)}>2/3$, we have $\Delta(\varrho)<0.075$, thus one can say that the two-quantum dots can be used for quantum teleportation.

\section{quantum thermodynamics}

In this section, we will investigate the extracted work of the two-quantum dots 1 and 2, as we have discussed between two-qubits \cite{amaziougqubit}. We proceeded through the schematics of the thermalization protocol phases with the assistance of a thermal bath, examined in \cite{Bellomo2019}. At time $t_1$, there are no interactions between the two-quantum dots. The interaction is then activated, and they thermalize simultaneously (from $t_2$ to $t_3$). These two-quantum dots are thermalized together from time $t_2$ to time $t_3$ ($t_3 - t_2 \gg \tau_r$, where $\tau_r$ represents the typical evolution time for the system at this phase). The extracted work can be evaluated via the partition functions of the system $Z_S$, ($\hbar = k_B = 1$) \cite{Bellomo2019}

\begin{equation} \label{H}
\mathcal{W} = T \ln\left( \frac{Z_1Z_2}{Z_S}\right) - \langle \mathcal{H}_{12} \rangle_{t_3},
\end{equation}
where $Z_{1(2)} = \Tr[\e^{\beta \mathcal{H}_{1(2)}}]$ is the partition function of system $1(2)$, $\langle \mathcal{H}_{12} \rangle_{t_3} = \Tr[\mathcal{H}\varrho (T)] - \Tr[(\mathcal{H}_1+\mathcal{H}_2)\varrho (T)]$, $\varrho (T)=\e^{-\beta \mathcal{H}}/Z_S$, $Z_S=\Tr[\e^{-\beta \mathcal{H}}]$ and $\beta=1/T$. The ideal efficiency is quantified as \cite{Bellomo2019}
\begin{equation}
\mu = \frac{\mathcal{W}}{\bigg(-\Delta\langle \mathcal{H}_{12} \rangle \bigg)}\leq 1
\end{equation}
where $\bigg (-\Delta \langle \mathcal{H}_{12} \rangle \bigg) = \langle \mathcal{H}_{12} \rangle_{t_2}-\langle \mathcal{H}_{12}\rangle_{t_3}$ with $\langle \mathcal{H}_{12} \rangle_{t_2}=0$, is the minimal work that must be done by the system for a single cycle to end. The global entropic term writes as $\mathcal{S}_G = T\bigg(\mathcal{S}(\varrho (T))-\mathcal{S}\left(\varrho_1(T)\otimes \varrho_2(T)\right)\bigg),$
where $\varrho_{1(2)}(T)=\e^{-\beta \mathcal{H}_{1(2)}}/Z_{1(2)}$ and $S(\hat\rho (T)) = - \Tr[\varrho (T) \ln(\varrho (T))]$. The energy difference is given by $\mathcal{E} = \mathcal{W} + \mathcal{S}_G,$
The local entropic term is writes as $\mathcal{S}_l = T \bigg (S(\varrho_1^{r} (T)\otimes \varrho_2^{r}(T)) - S(\varrho_1 (T)\otimes \varrho_2 (T))\bigg ),$
where $\varrho_{1(2)}^{r} (T) = \Tr_{2(1)}[\varrho (T)]$. The local work is given by $\mathcal{W}_l = \mathcal{W} - T\mathcal{S}(1:2),$, where $\mathcal{S}(1:2) = \mathcal{S}(\varrho_1^{r}(T)) + \mathcal{S}(\varrho_2^{r}(T)) - S(\varrho (T))$.

\begin{figure}[!htb]
\minipage{0.32\textwidth}
  \includegraphics[width=\linewidth]{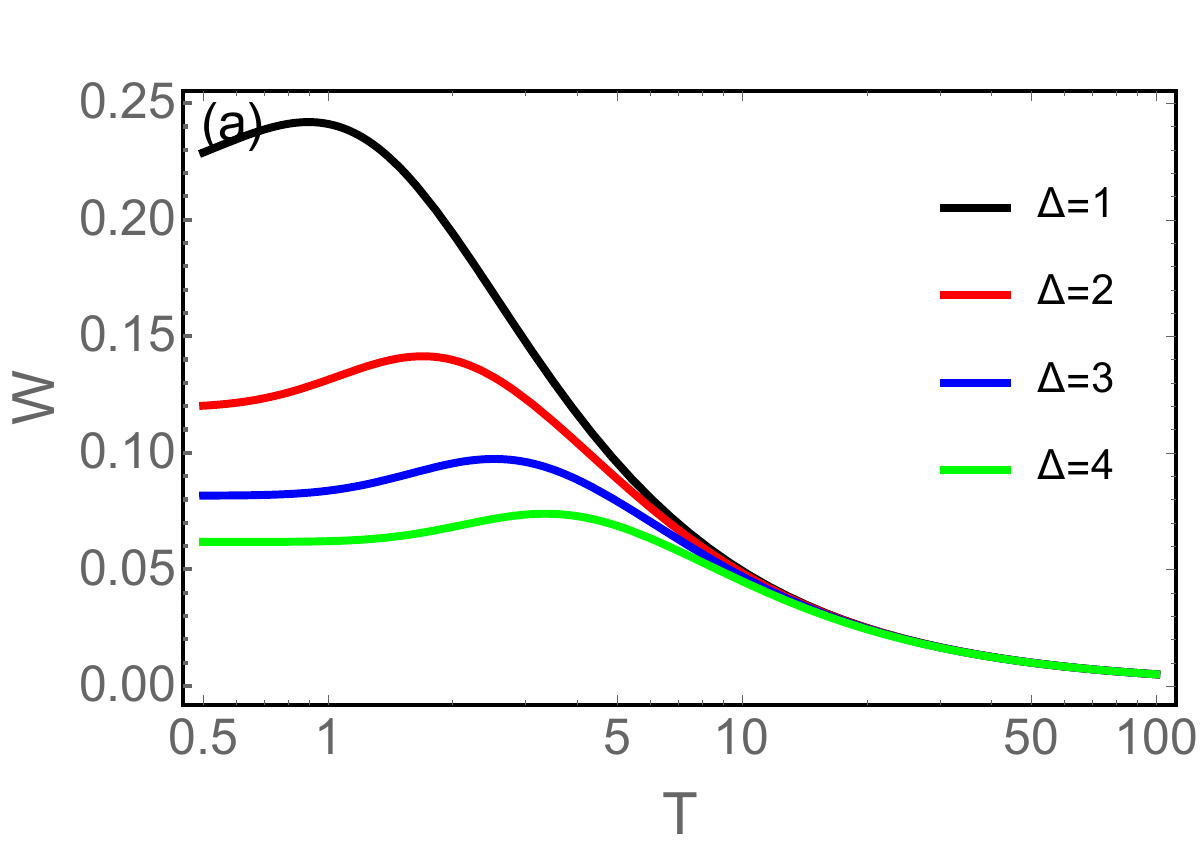}
\endminipage\hfill
\minipage{0.32\textwidth}
  \includegraphics[width=\linewidth]{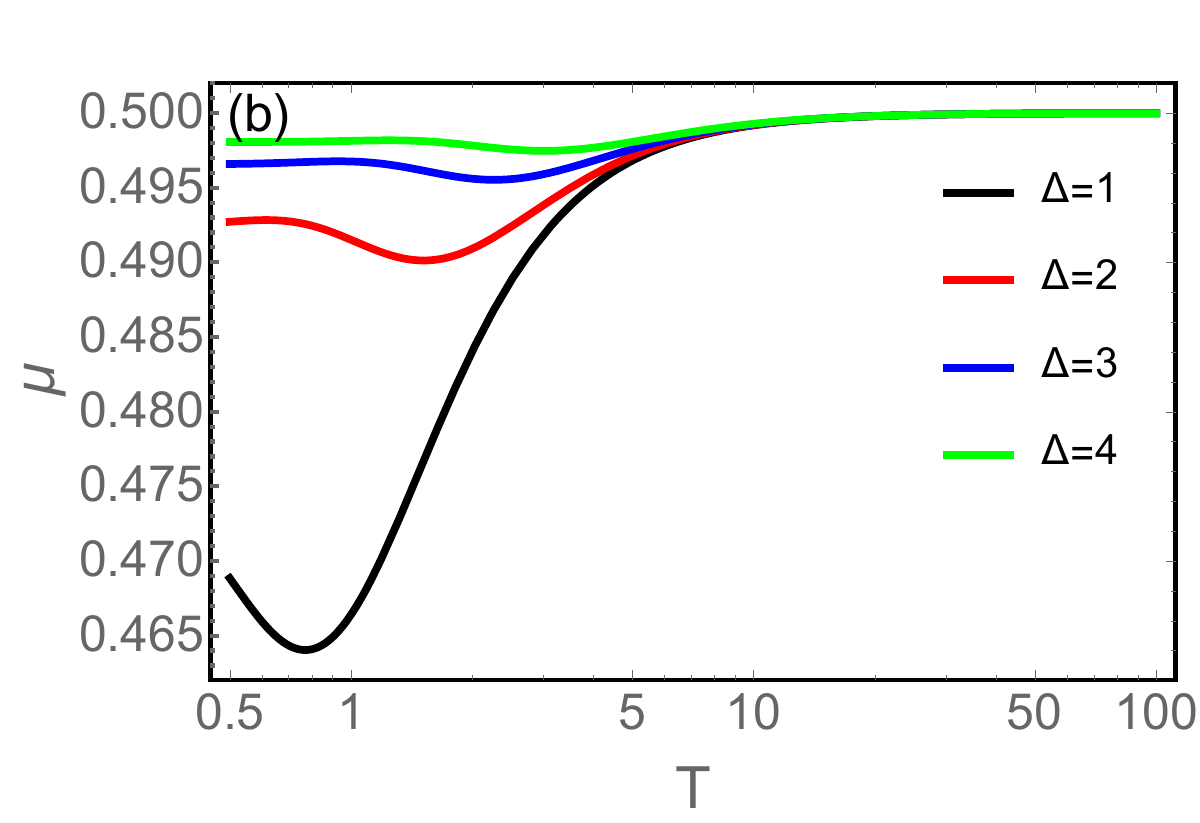}
\endminipage\hfill
\minipage{0.32\textwidth}
  \includegraphics[width=\linewidth]{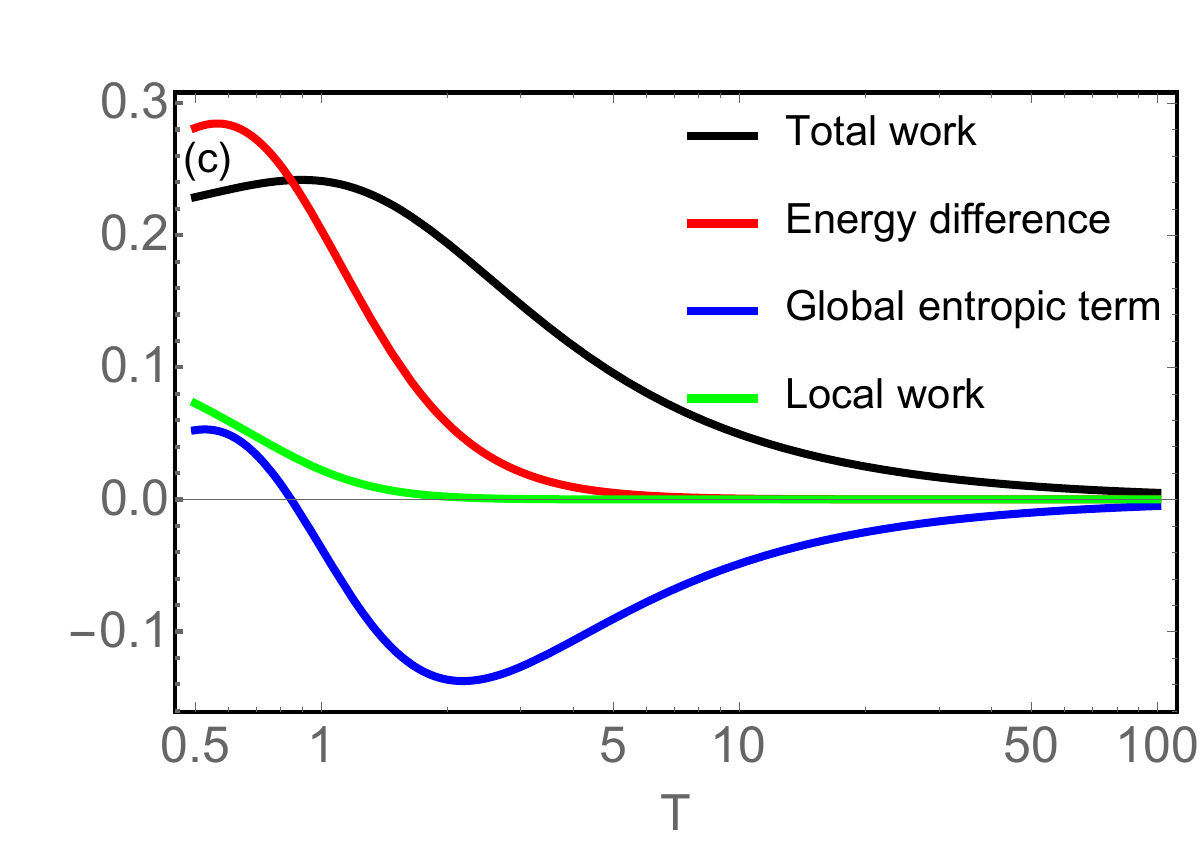}
\endminipage\hfill
\caption{Plot of (a) extracted work $\mathcal{W}$ and (b) efficiency $\mu$ of double quantum dot versus of the temperature $T$ for different values of the coupling tunneling, (c) comparison of different quantities for $J=1$.}
\label{W2}
\end{figure}

In Fig. \ref{W2}(a)-(b), we present the extracted work $\mathcal{W}$ and efficiency $\mu$ vs the bath temperature $T$ for different values of the tunneling coupling $\Delta$ of the two quantum dots. We note that the extracted work remains positive, and decreases after reaching its maximum value. The extracted work increases with increasing the tunneling coupling $\Delta$ for temperature less than 10 ($T<10$), as depicted in Fig. \ref{W2}(a). Moreover, when $T>10$, the influence of the coupling $\Delta$ on the extracting work is negligible. In Fig. \ref{W2}(b), exhibits that the efficiency ($\mu<1$) increases after achieves its minimum value to reaches its maximum value and to be constant around $\mu = 0.5$. Also, the efficiency increases with increasing the coupling tunneling $\Delta$, as implemented in Fig. \ref{W2}(b). these results are in direct agreement with respect to with present in \cite{Bellomo2019}.\\

We plot in Fig. \ref{W2}(c), different quantities such as the extracted work, energy difference, global entropy term and local work versus the temperature $T$, for $\Delta=1$ and $J=1$. We remark, that all quantities are positive for $T=0.5$ for small temperature, and tends to vanish for high temperature. According to this figure, the local work $\mathcal{W}_l$ decreases quickly with temperature to vanishes around $T=2$. Besides, the local work remains under total work, i.e., $W_l \leq W$. This means that the presence of correlations in the final thermal state benefits the amount of extracted work \cite{KMaruyama2009}. Regarding the global entropic term $\mathcal{S}_G$, is negative for a wide range of temperature, and is symmetric with respect to the total work $\mathcal{W}$, as shown in Fig. \ref{W2}(c). The total work $\mathcal{W}$ bounded by the energy difference $\mathcal{E}$ when $\mathcal{S}_G>0$. While when $\mathcal{S}_G<0$, the energy difference $\mathcal{E}$ diminishes quickly under total work $\mathcal{W}$ to vanishes.

\section{Conclusion} \label{Conc}

In summary, we have investigated and compared the quantum discord, entanglement, EPR-steering, CHSH-nonlocality, maximal fidelity and fidelity deviation of the two-quantum dots system under thermal effects and coupling tunneling. The quantum correlations beyond entanglement are quantified via the quantum discord. We have proved that the steerable state is strictly entangled but the entangled state is not always steerable. We have shown that quantum steering remains more persistent with respect to the quantum steering. We have found that the state of system under consideration can be used for quantum teleportation. We have evaluated the extracted work and efficiency. We have shown that the extracted work and efficiency can be enhanced by thermal effects. We compared, under thermal effect the total extracting work with differents energy quantities such as energy difference, global entropic term and local work. We have shown that the local work always bounded by total work.

\section*{Data availability statement}

No Data associated in the manuscript.

\end{document}